# Dual terahertz comb spectroscopy with a single free-running fibre laser

Guoqing Hu[1,2], Tatsuya Mizuguchi[1,3], Ryo Oe[1,3], Kazuki Nitta[1], Xin Zhao[2], Takeo Minamikawa[3,4], Ting Li[2], Zheng Zheng[2,5], and Takeshi Yasui[3,4]

[1]Graduate School of Advanced Technology and Science, Tokushima University, 2-1, Minami-Josanjima, Tokushima, Tokushima 770-8506, Japan

[2]School of Electronic and Information Engineering, Beihang University, Beijing, 100191, China

[3]JST, ERATO, MINOSHIMA Intelligent Optical Synthesizer Project, 2-1, Minami-Josanjima, Tokushima, Tokushima 770-8506, Japan

[4]Graduate School of Technology, Industrial and Social Sciences, Tokushima University, 2-1, Minami-Josanjima, Tokushima, Tokushima 770-8506, Japan

[5]Collaborative Innovation Centre of Geospatial Technology, 129 Luoyu Road, Wuhan 430079, China

Abstract

Dual THz comb spectroscopy has the potential to be used as universal THz spectroscopy with high spectral resolution, high spectral accuracy, and broad spectral coverage; however, the requirement for dual stabilized femtosecond lasers hampers its versatility due to the bulky size, high complexity, and high cost. We here report the first demonstration of dual THz comb spectroscopy using a single free-running fibre laser. By tuning the cavity-loss-dependent gain profile with an intracavity Lyot filter together with precise management of the cavity length and dispersion, dual-wavelength pulsed light beams with slightly detuned repetition frequencies are generated in a single laser cavity. Due to sharing of the same cavity, such pulsed light beams suffer from common-mode fluctuation of the repetition frequency, and hence the corresponding frequency difference between them is passively stable around a few hundred hertz within millihertz fluctuation. This considerably stable frequency difference enables dual THz comb spectroscopy with a single free-running fibre laser. While greatly reducing the size, complexity, and cost of the laser source by use of a single free-running fibre laser, the dual THz comb spectroscopy system maintains a spectral bandwidth and dynamic range of spectral power comparable to a system equipped with dual stabilized fibre lasers, and can be effectively applied to high-precision spectroscopy of acetonitrile gas at atmospheric pressure. The demonstrated results indicate that this system is an attractive solution for practical applications of not only THz spectroscopy but also THz-pulse-based measurements.

# Introduction

Optical spectroscopy is a powerful tool for material characterization in the solid, liquid, or gas phase. The interaction of a material with an optical wave varies depending on the wavelength, for example: electronic transitions for the visible light, intramolecular vibrations for near-infrared and mid-infrared light, intermolecular vibrations for far-infrared light or terahertz (THz) radiation, and plasma oscillations or orientational relaxation for microwaves. In particular, interesting interactions, such as hydrogen bonds or van der Waals interactions between molecules, phonon vibrations in crystal, or rotational vibrations of polar gas molecules, are still latent in the far-infrared or THz region (freq. = 0.1–10 THz, wavelength = 30–3000 µm, and photon energy = 0.37–41 meV) because the lack of coherent spectroscopic techniques in this region hampers the in-depth investigation of such interactions[1-3].

THz time-domain spectroscopy with broadband radiation (THz-TDS)[4] and THz frequency-domain spectroscopy with tunable narrow-linewidth radiation (THz-FDS)[5] have appeared as coherent spectroscopic techniques that are alternatives to conventional incoherent far-infrared spectroscopy[6] due to their unique characteristics, such as coherent radiation and detection or simultaneous acquisition of amplitude and phase spectra. Furthermore, THz-TDS also shows the possibility of being applied to pump–probe spectroscopy[7-9] or nonlinear spectroscopy[10,11]. However, these two methods cannot simultaneously achieve the high resolution, high accuracy, and wide bandwidth which are important performance metrics in spectroscopy.

THz frequency combs have attracted attention as precise frequency scales of broadband THz radiation in THz frequency metrology[12]. A THz comb possesses the characteristics of both broadband radiation and narrow-linewidth radiation due to its comb-tooth-like spectrum, and an absolute frequency of its all modes is secured by a frequency standard via a coherent frequency-comb link. Recently, the high potential of THz combs has been demonstrated in dual THz comb spectroscopy (THz-DCS)[13-20]. Based on multi-frequency heterodyning[14] or asynchronous optical sampling[13] of dual THz combs with different frequency spacings (= $f_{rep1}$, $f_{rep2}$), the frequency scale of THz combs has been downscaled by a ratio of $f_{rep1}/\Delta f_{rep}$ to the radio-frequency (RF) region as an RF comb with a frequency spacing of $\Delta f_{rep}$ (= $f_{rep2}$ - $f_{rep1}$), enabling us to obtain a mode-resolved THz comb spectrum. THz-DCS possesses the features of both THz-TDS and THz-FDS together with frequency traceability to a microwave or RF frequency standard, and hence can achieve high resolution, high accuracy, and broad bandwidth, simultaneously. The utility of THz-DCS has been demonstrated in high-precision broadband spectroscopy of polar molecular gas[16-20]. While THz-DCS has the potential to be used for universal THz spectroscopy, the use of a pair of dual femtosecond lasers and the requirement for stabilization control of $\Delta f_{rep}$ hamper the practical use of THz-DCS because of the difficulty of implementation and cost involved. Although adaptive sampling THz-DCS can eliminate the stabilization control requirement by cancelling the influence of the laser timing jitter[21], it still needs dual femtosecond lasers, which increases the system complexity. If THz-DCS can be implemented by a single free-running femtosecond laser, its versatility will be enhanced dramatically, and THz-DCS might become the *de facto* standard for THz spectroscopy in place of THz-TDS or THz-FDS.

In this article, we demonstrate THz-DCS by use of a single free-running dual-wavelength-comb (dual-λ-comb) Er-doped fibre (Er:fibre) laser. Two independent mode-locked oscillations in different wavelength regions are multiplexed in a single fibre cavity (upper row in Fig. 1)[22-24]. The dual-λ-comb light beams have different repetition frequencies, $f_{rep1}$ and $f_{rep2}$, due to wavelength dispersion of the cavity fibre, and their frequency spacing fluctuates in a common-mode manner due to sharing of the same cavity. Such common-mode fluctuation leads to high stability in $\Delta f_{rep}$ without the need for active laser stabilization. We generate dual THz combs with different frequency spacings by a combination of dual-λ-comb light beams and photoconductive antennae (middle row in Fig. 1), and apply the resulting RF comb (lower row in Fig. 1) to spectroscopy of molecular gas.

## **Results**

Figure 2 illustrates the experimental setup of the dual-λ-comb fibre laser and the THz-DCS system, which is described in the Methods section together with details of the experimental and analytical methodology employed for the following measurements.

First, we evaluated the basic performance of dual-λ-comb light radiating from the fibre laser oscillator. Figure 3a shows the optical spectra of the dual-λ-comb light: $\lambda_1$-comb light with a centre wavelength of 1533 nm and $\lambda_2$-comb light with a centre wavelength of 1543 nm. Although the spectral tails of the $\lambda_1$-comb light and $\lambda_2$-comb light were somewhat overlapped around 1537.5 nm, these comb light beams were independently mode-locked without coupling to each other. Evidence for two independent mode-locking oscillations was confirmed in the RF spectrum of the dual-λ-comb light (Fig. 3b): two frequency spikes appeared around 64.55 MHz, corresponding to different repetition frequencies of the dual-λ-comb light. Due to the low anomalous dispersion of the cavity fibre around this wavelength region, the $\lambda_1$-comb light had a higher repetition frequency $f_{rep1}$, whereas the $\lambda_2$-comb light had a lower repetition frequency $f_{rep2}$, and the frequency difference $\Delta f_{rep}$ between them was considerably small (< 400 Hz). We also evaluated the temporal fluctuations of $f_{rep1}$, $f_{rep2}$, and $\Delta f_{rep}$ because they determine the frequency-scale conversion factor (= $f_{rep1}/\Delta f_{rep}$) between the THz comb and the RF comb[14]. Figure 3c shows the temporal behaviour of the deviation in $f_{rep1}$ and $f_{rep2}$ ($\delta f_{rep1}$ and $\delta f_{rep2}$) from initial values with respect to the elapsed time. Due to the free-running or unstabilized operation, $f_{rep1}$ and $f_{rep2}$ showed a slow drift behaviour of several Hz within a range of 150 s. More importantly, these fluctuations were in common due to the common-path cavity. This resulted in millihertz fluctuation of $\Delta f_{rep}$ without drift behaviour, without active laser stabilization (Fig. 3d).

We next evaluated the basic performance of the present THz-DCS system equipped with a single free-running dual-λ-comb Er:fibre laser ($f_{rep1} \approx f_{rep2} \approx$ 64.55 MHz, $\Delta f_{rep}$ = 248.4 Hz), namely, a single-free-running THz-DCS system, by comparing it with our previous THz-DCS system equipped with dual stabilized Er:fibre lasers ($f_{rep1} \approx f_{rep2} \approx$ 250.00 MHz, $\Delta f_{rep}$ = 893.00 Hz, see Methods)[17-20], namely, a dual-stabilized THz-DCS system. Both THz-DCS systems used the same THz optical systems and data acquisition electronics. Figure 4a shows a comparison of the THz power spectrum between the single-free-running and dual-stabilized THz-DCS systems when the data acquisition time was set to 100 s. The spectral resolution was set to 1 GHz. For reference, the noise spectrum is shown in the same graph. The spectral bandwidth and dynamic range of spectral power were comparable to each other. We also investigated the relation between the data acquisition time and dynamic range of spectral power within a frequency range of 0.2 to 0.4 THz (see Fig. 4b). The linear relationship between them clearly indicated that the timing jitter in both systems had little influence. Although the dynamic range in the single-free-running THz-DCS system was higher than that in the dual-stabilized THz-DCS systems, this difference was mainly due to difference of repetition frequencies between them. If such difference is corrected, both were almost overlapped (see Fig. 4c). In this way, the passive stabilization of $\Delta f_{rep}$ in the free-running dual-λ-comb Er:fibre laser was as powerful as the active stabilization of $f_{rep1}$ and $f_{rep2}$ in the dual stabilized Er:fibre lasers.

Finally, we demonstrated THz spectroscopy of acetonitrile ($CH_3CN$) gas under atmospheric pressure. Since $CH_3CN$ gas is contained in the interstellar medium[1], incomplete combustion gas of nylon textiles[25], volatile organic compounds related with atmosphereic pollution and biomarkers[26], it is important to perform THz spectroscopy of this molecular gas in astronomy, fire accidents, atmospheric analysis, and health monitoring. Due to the symmetric top molecule, $CH_3CN$ gas exhibits characteristic spectral fingerprints with GHz structure in the THz region under atmospheric pressure: a series of manifolds of multiple rotational transitions regularly spaced by *2B*, where *B* is the rotational constant (= 9.194 GHz)[27]. After an enclosed box for THz optics was filled with $CH_3CN$ gas at atmospheric pressure, the THz power spectrum was acquired by the single-free-running THz-DCS system. Figure 5a shows the absorbance spectrum of $CH_3CN$ gas ($f_{rep1} \approx f_{rep2} \approx$ 64.55 MHz, $\Delta f_{rep}$ = 388.6 Hz, spectral resolution = 1 GHz, data acquisition time = 257 s). Thirty-nine manifolds of absorption lines periodically appeared with a constant frequency separation, and could be assigned to rotational quantum numbers from *J*=15 around 0.29 THz to *J*=53 around 0.98 THz correctly. Figure 5b shows the magnified absorbance spectrum of the same $CH_3CN$ gas within a frequency range of 0.3 to 0.4 THz. For reference, the literature values of integrated intensity for $CH_3CN$ gas in the JPL spectral database[27] is shown as purple lines in Fig. 5b. Although neighbouring manifolds of rotational transitions were somewhat overlapped due to the pressure broadening at atmospheric pressure, the spectra show similar signatures to the literature values, namely, a periodic structure of absorption peaks with a constant spacing exactly equal to *2B* (= 18.4 GHz). In this way, we confirmed the effectiveness of the single-free-running THz-DCS for atmospheric-pressure gas spectroscopy in the THz region.

## Discussion

We confirmed that use of the single free-running fibre laser did not degrade the spectroscopic performance of THz-DCS, and that the single-free-running THz-DCS system has the potential for atmospheric-pressure gas spectroscopy. Here, we first discuss the limitations of spectral resolution and accuracy. The spectra resolution was limited to 1 GHz due to the time window size of 1 ns in the temporal waveform of the pulsed THz radiation. On the other hand, the instability of the conversion factor limits the spectral accuracy because the frequency scale of the THz comb was decreased to the RF region based on a conversion factor of $f_{rep1}/\Delta f_{rep}$. Figure 6a shows the frequency instability of $f_{rep1}$ and $\Delta f_{rep}$ for the single free-running fibre laser. The frequency instability was defined as the ratio of the standard deviation to the mean of $f_{rep}$ and $\Delta f_{rep}$. The instability of the conversion factor is essentially limited by that of $\Delta f_{rep}$ rather than $f_{rep1}$, and was achieved $10^{-4}$ in the data acquisition time of a few hundred seconds. This value directly corresponds to the spectral accuracy in THz-DCS. On the other hand, Fig. 6b show the frequency instability of $f_{rep1}$ and $\Delta f_{rep}$ for the dual stabilized Er:fibre lasers, indicating that the instability of the conversion factor was achieved to $10^{-8}$ in the data acquisition time of a few hundred seconds. The dual-stabilized THz-DCS showed a 10,000-times better spectral accuracy than single-free-running THz-DCS; however, the difference in spectral accuracy between them was negligible in atmospheric-pressure gas spectroscopy due to the pressure-broadening linewidth of several gigahertz to a few tens of gigahertz. A combination of single-free-running THz-DCS and the adaptive sampling method[21] will enable us to perform high-precision THz spectroscopy equal to or greater than dual-stabilized THz-DCS system by powerful correction of the laser timing jitter. It is also interesting to compare the frequency stability of $f_{rep1}$ and $\Delta f_{rep}$ between the single free-running laser and dual free-running lasers. To this end, we evaluated the frequency instability of them when dual stabilized lasers were operated without the frequency control, namely dual free-running lasers (see Fig. 6b). The $f_{rep1}$ instability of the dual free-running laser was comparable to that of the single free-running lasers (see Fig. 6a). However, the $f_{rep1}$ instability of the dual free-running laser was significantly worse than that of the single free-running lasers, clearly indicating the advantage to sharing of the same cavity.

Next, we discuss the practicability of the single-free-running THz-DCS system. Compared with the dual-stabilized THz-DCS system, use of the single free-running fibre laser largely enhances the practicability from the viewpoints of compactness, cost effectiveness, and ease of implementation of the laser source. An inset of Fig. 2 shows the optical photograph of the single free-running dual-λ-comb

Er:fibre laser, including a single laser oscillator (size: 450 mm width × 350 mm depth × 100 mm height), EDFAs (size: 150 mm width × 120 mm depth × 20 mm height) and a LD driver (size: 100 mm width × 150 mm depth × 50 mm height). On the other hand, dual stabilized Er:fibre lasers are composed of dual laser heads (size: 415 mm width × 400 mm depth × 110 mm height) and control electronics (mounted in a 19" rack cabinet, size: 600 mm width × 800 mm depth × 1800 mm height). Although the size of the dual-λ-comb Er:fibre laser is comparable to that of dual stabilized Er:fibre lasers, there are still space to largely reduce its size in the former due to the simple cavity configuration without the need for laser control (see Fig. 2). Regarding the accompanying electronics, the total volume of the dual-λ-comb Er:fibre laser was reduced to 0.086 % of the volume of the dual stabilized Er:fibre lasers. Such a large reduction of the laser size will lead to the development of portable THz-DCS systems with flexible and robust fibre-coupled photoconductive antennae. Also, the lack of need for laser stabilization electronics drastically decreases the cost and complexity of the laser source as well as its size. In this way, despite the fact that the single-free-running THz-DCS system uses a much more compact, more cost-effective, and more easy-to-operate laser than the dual-stabilized THz-DCS system, its spectroscopic performance is comparable to that of the previous THz-DCS system.

In conclusion, we greatly enhanced the versatility of THz-DCS by use of a single free-running dual-λ-comb Er:fibre laser without degrading its high performance

- 11 -

in terms of spectral resolution, spectral accuracy, and broad spectral bandwidth. By using a cavity configuration similar to that of usual mode-locked fibre lasers, dual-λ-comb light beams with passively stable $\Delta f_{rep}$ were generated via wavelength-region multiplexing of mode-locked oscillations in the same cavity. Use of such dual-λ-comb light beams in THz-DCS will enable a drastic reduction in size, complexity, and cost. Although versatility always conflicts with high performance, this dual-λ-comb Er:fibre laser enable us to balance these competing characteristics in broadband high-precision THz spectroscopy, which is the first time to the best of our knowledge. This versatile and high-performance THz-DCS will be a powerful tool as a simple, portable, universal, fast THz spectrometer for gas analysis[20], biosensor[28], impulse radar[29], and so on.

## Methods

*THz-DCS*

A THz comb spectrum can be obtained by the multi-frequency-heterodyning method in the frequency domain (Fig. 2)[14] or the asynchronous-optical-sampling (ASOPS) method in the time domain[13,17]. We here used the latter method. Supplementary Fig. 1 shows a signal flowchart of THz-DCS used here. The ASOPS method enabled us to linearly magnify the time scale of a THz pulse train having a repetition period of $1/f_{rep1}$ based on a temporal magnification factor $f_{rep1}/\Delta f_{rep}$ to give an RF pulse train with a repetition period of $1/\Delta f_{rep}$. The RF pulse train could be directly acquired by a digitizer without the need for mechanical time-delay scanning. The Fourier transform of this RF pulse train gave an RF comb spectrum with a frequency spacing of $\Delta f_{rep}$. Finally, frequency calibration of the RF comb with the temporal calibration factor enabled us to obtain a THz comb spectrum with a frequency spacing of $f_{rep1}$.

*Dual-λ-comb Er:fibre laser*

The dual-λ-comb fibre laser oscillator had a ring cavity including a cavity single-mode fibre (SMF1; SMF28, Corning, dispersion = 18 ps/nm/km at 1550 nm, length = 1.79 m), another cavity single-mode fibre (SMF2; HI1080, Corning, dispersion = -53 ps/nm/km at 980 nm, length = 0.11 m), a cavity erbium-doped fibre (EDF; ER110-4/125, LIEKKI, dispersion = -9.6 ps/nm/km at 1500-1600 nm, length = 0.56 m), a cavity dispersion compensation fibre (DCF; DCF 180, Yangtze Optical Fibre and Cable Co., Ltd., dispersion = 180 ps/nm/km at 1550 nm, length = 0.15 m), a single-wall carbon nanotube (SW-CNT) mode-locker, an in-line polarizer with polarization-maintained fibre pigtails (ILPL-PMF), a polarization controller (PC), an wavelength division multiplex (WDM), an output coupler (OC), and a pump laser diode (LD, wavelength = 980 nm). The ILPL-PMF functioned as a Lyot filter in the cavity. Dual-λ mode-locking oscillation was achieved by adjusting the gain profile of the EDF and the birefringence-induced spectral filtering with the pump power and PC. Our previous dual-λ-comb Er:fibre laser provided a stable $\Delta f_{rep}$ around 1.76 kHz[24]; however, this $\Delta f_{rep}$ value is too high for THz-DCS because the high $\Delta f_{rep}$ value leads to a low temporal magnification factor (= $f_{rep1}/\Delta f_{rep}$). We reduced the total dispersion to 7 fs/nm/km in a cavity length of 3.2 m by a combination of SMF, EDF, and DCF, and achieved an $\Delta f_{rep}$ value much lower than that of the previous laser. Output light beams from the dual-λ-comb fibre laser oscillator were separated into short-wavelength $\lambda_1$-comb light (centre wavelength = 1533 nm) and long-wavelength $\lambda_2$-comb light (centre wavelength = 1543 nm) by a coarse-wavelength-division-multiplexing bandpass filter (CWDM-BPF, passband = 1530±7.5 nm). Supplementary Figure 2a shows optical spectra of the spectrally separated $\lambda_1$-comb light and $\lambda_2$-comb light. The resulting $\lambda_1$-comb light and $\lambda_2$-comb light were amplified and spectrally broadened by a pair of Er:fibre amplifiers (EDFAs), and then their pulse duration was minimized by dispersion control with SMF and DCF. After passing through an isolator (ISO) and OC, the $\lambda_1$-comb light (centre wavelength = 1531 nm, mean power = 20 mW, and pulse duration = 130 fs) was used for probe light, whereas the $\lambda_2$-comb light (centre wavelength = 1543 nm, mean power = 27 mW, and pulse duration = 130 fs) was used for pump light in the THz-DCS system. Supplementary Fig. 2b shows optical spectra of the amplified $\lambda_1$-comb light and $\lambda_2$-comb light, and Supplementary Figs. 2c and 2d show their auto-correlation trace.

*Dual stabilized Er:fibre lasers*

Dual $f_{rep}$-stabilized, mode-locked Er-fibre lasers (ASOPS TWIN 250, Menlo Systems, centre wavelength = 1550 nm, pulse duration = 50 fs, $f_{rep1} \approx f_{rep2} \approx$ 250 MHz) were used in the dual-stabilized THz-DCS system for comparison with the single-free-running THz-DCS system. The frequencies $f_{rep1}$ and $f_{rep2}$ were stabilized at 250,000,000 Hz and 250,000,893 Hz by two independent laser control systems (RMS timing jitter < 150 fs in the range 0.1 Hz – 500 kHz) referenced to a rubidium frequency standard (Rb-FS; FS725, Stanford Research Systems, accuracy = 5×10$^{-11}$ and instability = 2×10$^{-11}$ at 1 s). Thus, the frequency difference between them ($\Delta f_{rep}$ = $f_{rep2}$ – $f_{rep1}$ = 893 Hz) was also stabilized. These values of $f_{rep1}$ and $\Delta f_{rep}$ give the same TMF as the single-free-running THz-DCS system, enabling us to use the same data acquisition electronics in both systems. The $f_{rep1}$ light was used for pump light, whereas the $f_{rep2}$ light was used for probe light in THz-DCS system.

*Experimental setup*

A pair of fibre-coupled LT-InGaAs/InAlAs photoconductive antennas (PCAs) was used for the THz-DCS system: a strip-line-shaped LT-InGaAs/InAlAs PCA (PCA1, TERA 15-TX-FC, Menlo Systems, bias voltage = 20 V, optical power = 20 mW) for a THz emitter and a dipole-shaped LT-InGaAs/InAlAs PCA (PCA1, TERA 15-RX-FC, Menlo Systems, optical power = 20 mW) for a THz detector. THz pulse train was radiated from the PCA1 triggered by the $\lambda_2$-comb light (pump light), propagated in an enclosed box, and was then incident on PCA2 together with the $\lambda_1$-comb light (probe light). The enclosed box was filled with nitrogen gas or acetonitrile gas. The temporal waveform of the output current signal from PCA2 was acquired as the RF pulse train by a digitizer (National Instruments, PCI-5122, sampling rate = 1×10$^8$ samples/s, resolution = 14 bit) after amplification with a current preamplifier (AMP; FEMTO

Messtechnik GmbH, HCA-10M-100K, bandwidth = 10 MHz, trans-impedance gain = $1\times10^{5}$ V/A). Portions of the $\lambda_1$-comb light and the $\lambda_2$-comb light separated by OCs were fed into a sum-frequency-generation cross-correlator (SFG-X) for generation of a trigger signal for the digitizer. A rubidium frequency standard (FS725, Stanford Research Systems, accuracy = $5\times10^{-11}$ and instability = $2\times10^{-11}$ at 1 s) was used for an external clock signal in the digitizer. The sampling interval and time window size were set to 60 fs and 15.5 ns, respectively, for the single-free-running THz-DCS system and to 14 fs and 4 ns, respectively, for the dual-stabilized THz-DCS system.

*Data analysis*

We set the spectral resolution of the THz-DCS system to be 1 GHz to fit its spectral resolution with the pressure-broadening absorption linewidth. To this end, a temporal waveform of the pulsed THz radiation was extracted with a time window size of 1 ns from the whole temporal waveform of the pulsed THz radiation. Then, the THz power spectrum was obtained by taking the Fourier transform of the temporal waveform and subsequent squaring, and the result was used for calculation of the absorbance spectrum. The absorbance spectrum was obtained by using the THz power spectrum obtained when the enclosed box was filled with nitrogen gas as a reference.

*Data availability*

The data that support the findings of this study are available from the corresponding author upon reasonable request.

**Acknowledgements**

The work at Tokushima University was supported by grants for the Exploratory Research for Advanced Technology (ERATO) MINOSHIMA Intelligent Optical Synthesizer (IOS) Project (JPMJER1304) from the Japanese Science and Technology Agency and a Grant-in-Aid for Scientific Research (A) No. 26246031 from the Ministry of Education, Culture, Sports, Science, and Technology of Japan. The work at Beihang University was supported by the 973 Program (2012CB315601), NSFC (61521091/61435002) and with Fundamental Research Funds for the Central Universities, Beihang PhD Student Funds for Short-term Visiting Study and the Academic Excellence Foundation of BUAA for PhD Students.

**Author contributions**

T. Y. and Z. Z. conceived the project. G. H., Tat. Miz., R. O., and K. N. performed the experiments and/or analysed the data. X. Z. and T. L. contributed to the dual-comb sources. G. H. and T. Y. wrote the manuscript. Tak Min. discussed the results and commented on the manuscript.

**Competing financial interests statement**

The authors declare no competing financial interests.

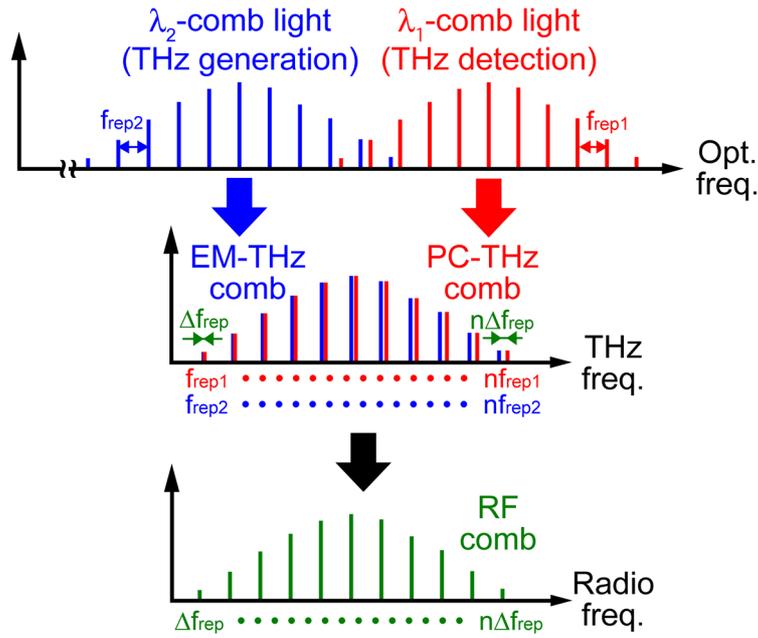

**Fig. 1. Principle of dual THz comb spectroscopy.** $\lambda_1$-comb light (frequency spacing = $f_{rep1}$) is incident on a photoconductive antenna for THz detection (PCA detector), resulting in generation of a photocarrier THz comb (PC-THz comb, frequency spacing = $f_{rep1}$) in the PCA detector. $\lambda_2$-comb light (frequency spacing = $f_{rep2}$) is incident on a photoconductive antenna for THz generation (PCA emitter), resulting in radiation of an electromagnetic THz comb (EM-THz comb, frequency spacing = $f_{rep2}$) from the PCA emitter. When the free-space-propagating EM-THz comb is detected by the PCA detector having the PC-THz comb, a secondary frequency comb in the RF region (RF comb, frequency spacing = $\Delta f_{rep} = f_{rep1} - f_{rep2}$) is generated as a current signal from the PCA detector via multi-frequency-heterodyning photoconductive detection between the EM-THz comb and PC-THz comb. The RF comb is a replica of the EM-THz comb whose frequency spacing is downscaled from $f_{rep2}$ to $\Delta f_{rep}$ by a conversion factor $f_{rep2}/\Delta f_{rep}$.

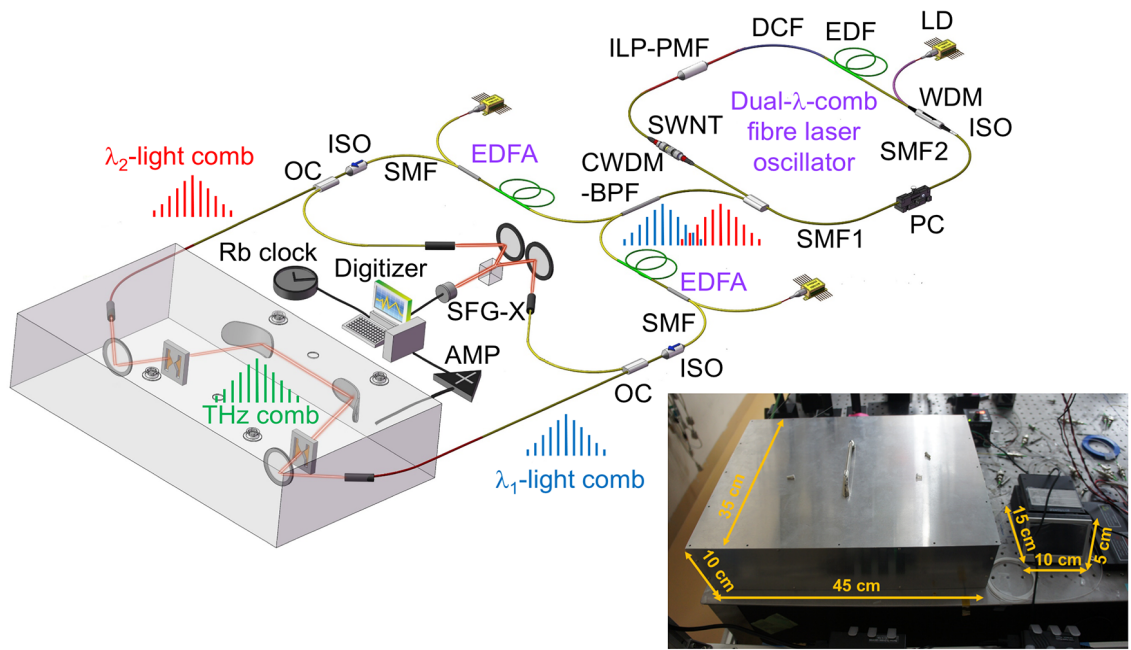

Fig. 2. **Experimental setup.** An inset shows an optical photograph of dual-λ-comb fibre laser oscillator. See Methods for details.

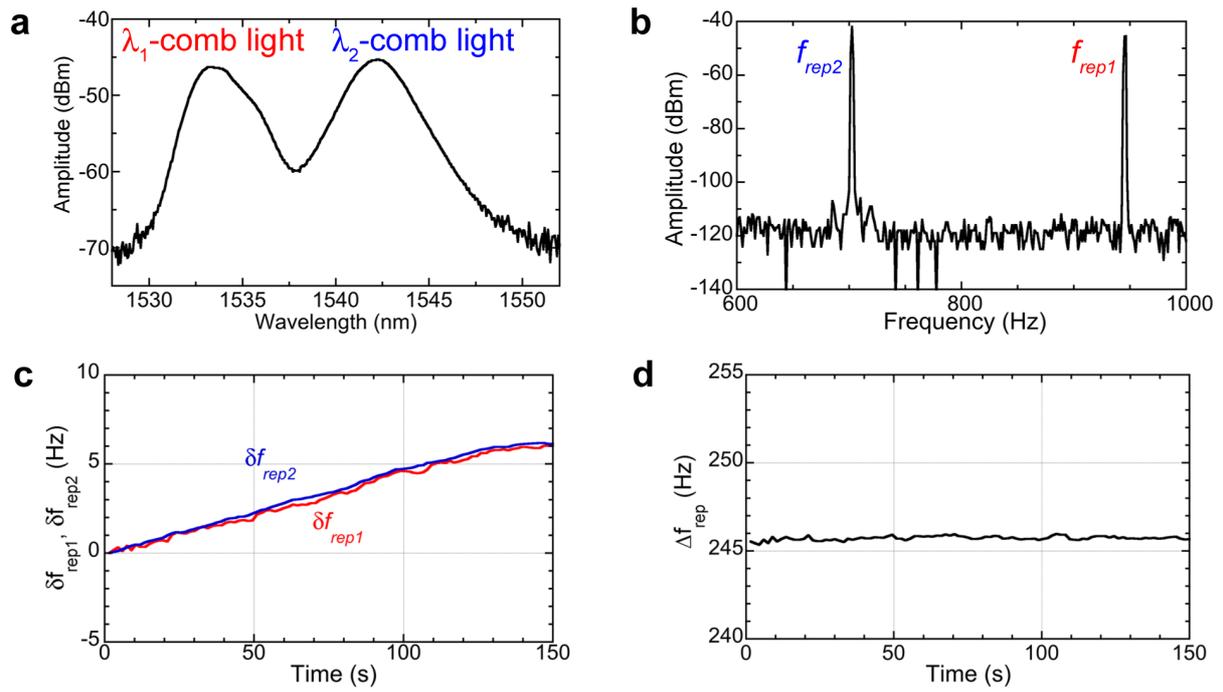

**Fig. 3. Basic performance of dual-λ-comb Er:fibre laser oscillator light. a**, Optical spectrum of dual-λ-comb light. **b**, RF spectra of repetition frequency signal in dual-λ-comb light. **c**, Temporal fluctuations of *δf$_{rep1}$* and *δf$_{rep2}$*. **d**, Temporal fluctuations of Δ*f$_{rep}$*.

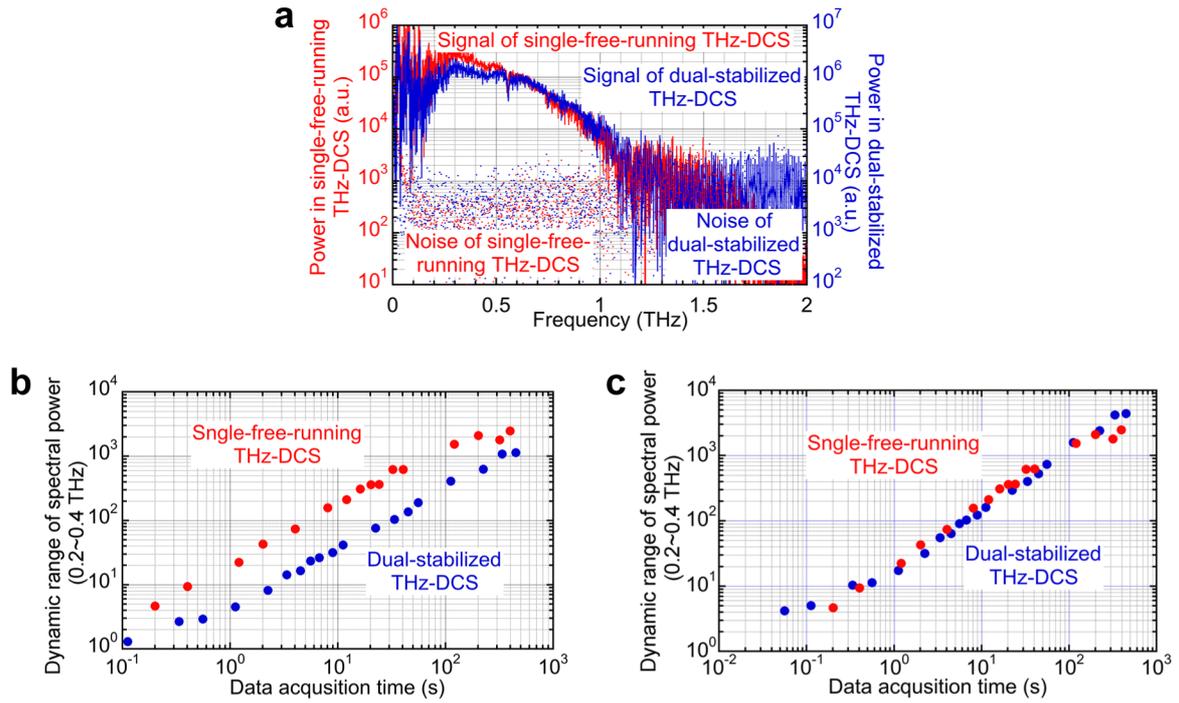

Fig. 4. **Comparison of basic performance between single-free-running and dual-stabilized THz-DCS systems. a**, Spectral bandwidth. Dynamic range of THz spectral power **b**, before and **c**, after correction of different repetition frequencies.

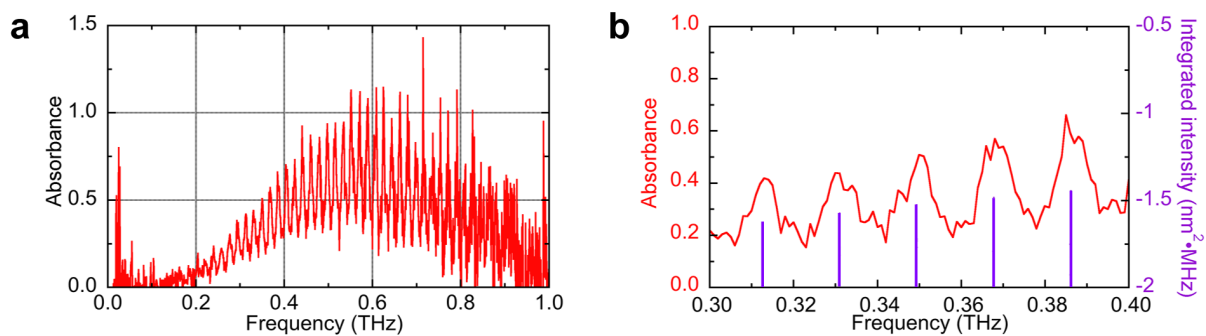

Fig. 5. **Spectroscopy of $CH_3CN$ gas in the atmospheric pressure. a**, Absorbance spectrum of $CH_3CN$ gas within a frequency range of 0 to 1 THz. **b**, Magnified absorbance spectrum of $CH_3CN$ gas within a frequency range of 0.3 to 0.4 THz. Purple lines show literature values of integrated intensity for $CH_3CN$ gas in the JPL database [ref@5e].

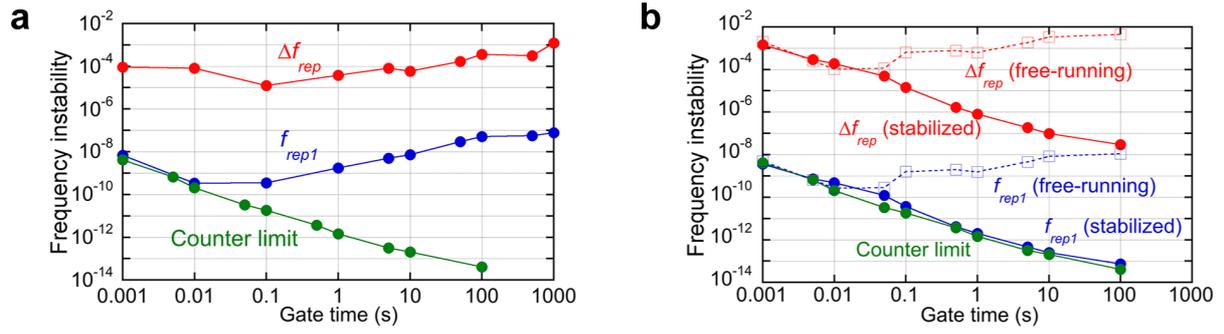

Fig. 6. **Frequency instability in $f_{rep1}$ and $\Delta f_{rep}$ with respect to gate time. a**, Single-free-running THz-DCS. **b**, Dual-stabilized THz-DCS.

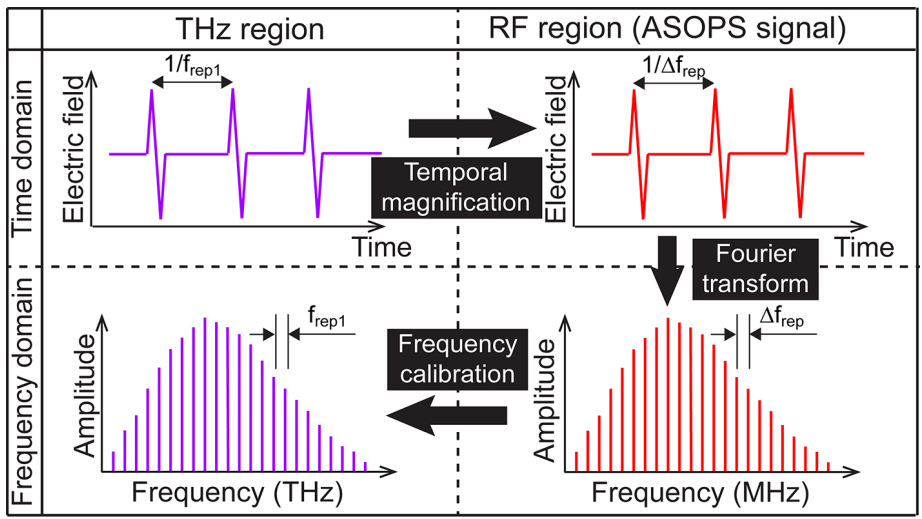

Supplementary Fig. 1. **Principle of THz-DCS in time domain.** See Methods for details.

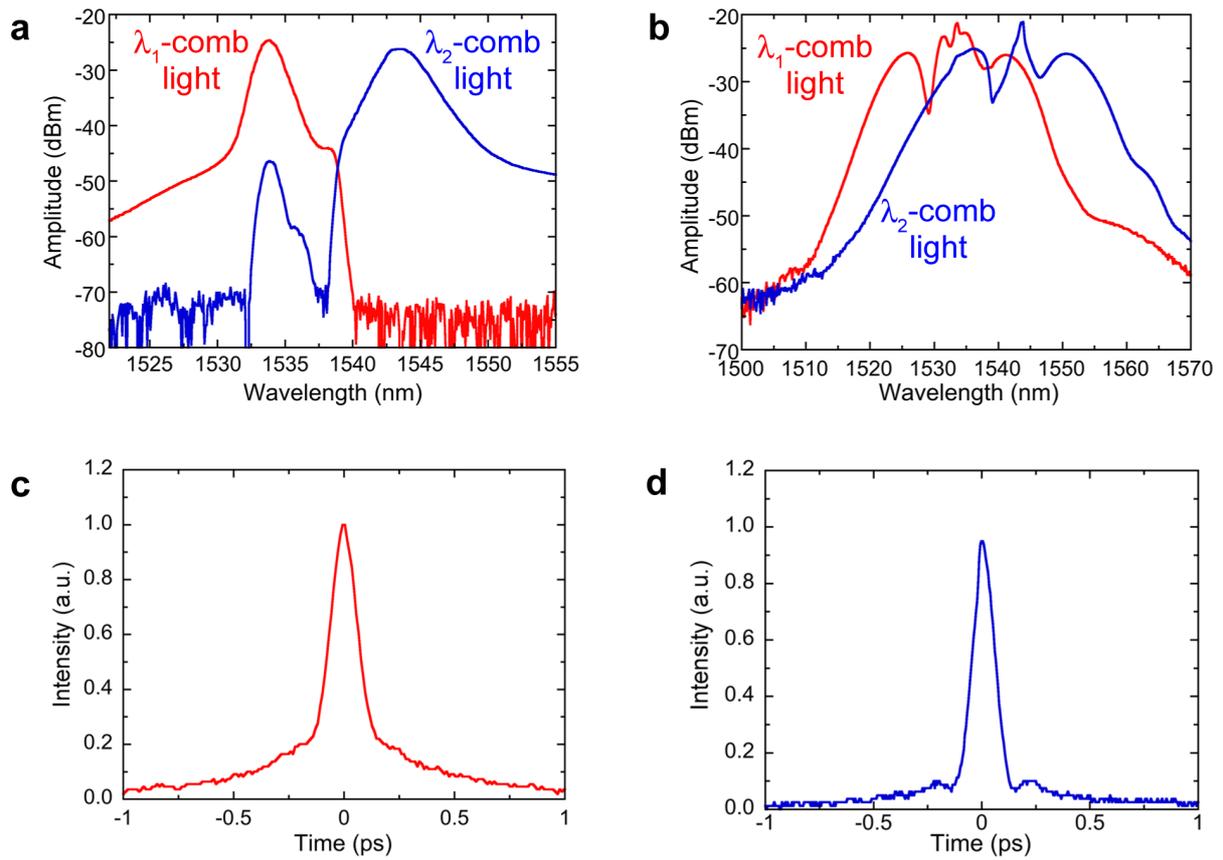

Supplementary Fig. 2. **Basic performance of amplified dual-λ-comb Er:fibre laser light. a**, Optical spectra of separated $\lambda_1$-comb light and $\lambda_2$-comb light. **b**, Optical spectra of amplified $\lambda_1$-comb light and $\lambda_2$-comb light. **c**, Auto-correlation trace of amplified $\lambda_1$-comb light. **d**, Auto-correlation trace of amplified $\lambda_2$-comb light.